\documentclass[preprint,showpacs,preprintnumbers,amsmath,amssymb]{revtex4}

\usepackage{graphicx}
\usepackage{dcolumn}
\usepackage{latexsym}
\usepackage{amsmath}
\usepackage{amssymb}
\usepackage{bm}
\usepackage{textcomp}

\begin{document}

\preprint{APS/123-QED}

\title{Impact of Electrostatic Forces in Contact Mode Scanning Force Microscopy}

\author{F. Johann\footnote{Present address: Max Planck Institute for Microstructure Physics, Weinberg 2, 06120 Halle, Germany}}
\author{\'{A}. Hoffmann}
\author{E. Soergel}
\email{soergel@uni-bonn.de}

\affiliation{Institute of Physics, University of Bonn,
Wegelerstra\ss e 8, 53115 Bonn, Germany}


\begin{abstract}
In this $\ll$ contribution we address the question to what extent surface charges affect contact-mode scanning force microscopy measurements.
We therefore designed samples where we could generate localized electric field distributions near the surface as and when required.
We performed a series of experiments where we varied the load of the tip, the stiffness of the cantilever and the hardness of the sample surface.
It turned out that only for soft cantilevers could an electrostatic interaction between tip and surface charges be detected, irrespective of the surface properties, i.\,e. basically regardless its hardness. %
We explain these results through a model based on the alteration of the tip-sample potential by the additional electric field between charged tip and surface charges.
\end{abstract}

\pacs{68.37.Ps, 77.84.-s, 77.80.Dj}

\maketitle

\section{Introduction} \label{intro}
The omnipresence of scanning probe microscopes is because of their versatility for the characterization of different surface properties, some of them with a lateral resolution down to the sub-nm regime and an otherwise unequaled sensitivity~\cite{MHB,Sarid}. Scanning force microscopes in particular are often used to map the topography and simultaneously record the spatial distribution of, e.\,g.,~the magnetization~\cite{Rug90}, the charge mobility~\cite{Dai96}, or the surface potential~\cite{Fuj92}. A very prominent example of this  technique is electrostatic force microscopy (EFM)~\cite{Ste88}, a method for mapping surface charge distributions with a sensitivity allowing for single charge detection~\cite{Sch90}. In brief, a scanning force microscope (SFM) is operated in non-contact mode with an additional oscillating voltage applied to the tip. The electrostatic interaction between the periodically charged tip and the charges at the sample surface to be measured leads to a vibration of the cantilever that can be read-out with a lock-in amplifier. Interestingly, a very similar technique is used in piezoresponse force microscopy (PFM)~\cite{Alexe}. Here the SFM is operated in contact mode, and the piezomechanical response of the sample, i.\,e.~its periodic thickness changes caused by the applied alternating voltage, is measured again with a lock-in amplifier. This latter technique is mainly applied to the detection of ferroelectric domains as ferroelectricity entails piezoelectricity. Now all ferroelectric materials show by nature a domain-specific surface polarization charging. Thus the question about a possible contribution of electrostatic forces between the charged tip and the polarization surface charges to the measured PFM signals arises. In other words: is an SFM operated in contact mode capable of mapping surface charge distributions?

Our aim is to investigate this subject thoroughly as in several publications using piezoresponse force microscopy an electrostatic contribution to the measured signals is taken for granted~\cite{Fra95,Fra98,Lee99,Hon01,Har02,Kal02,Hon02,Rod02,Ros03,Sva05,Scr05,Set06,Fra06,Gru06}.
Although this presumed electrostatic contribution is not quantified,
it is assumed to influence significantly the measured signals in PFM. In  a couple of publications the authors even attribute the contrast in contact mode SFM imaging of ferroelectric domains fully to the electrostatic interaction between the charged tip and the polarization charges at the sample's surface~\cite{Hon98,Hon99,Luo00,Shi04,Patent}. In order to obtain quantitatively reliable results from PFM-measurements, it is therefore  mandatory to quantify a possible electrostatic contribution to the measured PFM signals.

In this contribution, we present a series of measurements with especially designed samples to clarify whether there is any, and in any case to quantify the electrostatic contribution in contact mode scanning  force microscopy measurements. To determine the influence of the sample properties, i.\,e.~mainly it's hardness, we also prepared a series of samples out of different materials. Furthermore, we used different cantilevers and altered the load of the tip. In addition to our experimental investigations we propose a model to explain the observed dependencies of the measured signals on the tip load, the cantilever stiffness and the samples' material properties.

\section{Basic Considerations} \label{sec:basic}
In scanning force microscopy a tiny tip attached to the end of a cantilever is scanned across the surface. A displacement of the tip and/or the cantilever results in a deformation of the latter that can be measured by, e.\,g.,~the beam deflection method, a so called 'SFM-signal'. A SFM-signal caused by electrostatic interaction between the probe and the sample will be named 'electrostatically caused SFM-signal' (SFM$_{\rm el}$-signal).

\begin{figure}[ppp]
\includegraphics{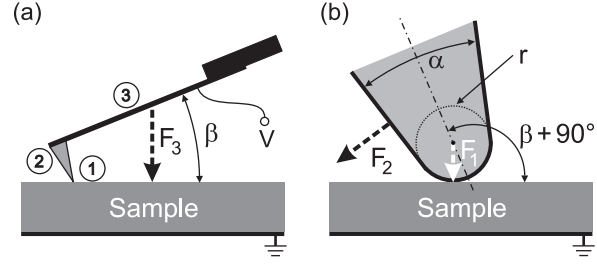}
\caption{\label{fig:Joh01}
(a)~Summary of the possible electrostatic interactions between the
sample surface and the probe with a voltage $V$ applied to it. The
probe consists of \textcircled{1} sphere, \textcircled{2} cone, and
\textcircled{3} cantilever. $F_3$~denotes the electrostatic force
between the cantilever and the sample surface. The angle between
cantilever and sample surface is $\beta$.
(b)~Close view of the tip, i.\,e.~a sphere of radius~$r$ and the
cone (opening angle~$\alpha$). $F_2$~denotes the electrostatic force
between the cone and the sample surface. $F_1$~describes the electrostatic force between the the sphere and the sample surface.
}
\end{figure}

In the following, we will discuss separately the possible mechanisms
for the tip (i.\,e.~sphere and cone) and the cantilever to contribute to an SFM$_{\rm el}$-signal when operating the SFM in contact-mode, i.e.~the apex of the tip is located in the repulsive part of the Lennard-Jones potential between tip and sample surface. An overview of the situation is shown in~Fig.~\ref{fig:Joh01}. For our considerations, we assume the probe to be conductive and electrically connected to an oscillating voltage source (amplitude: some volts, frequency: several 10\,kHz).


\subsection{Tip-Sample Interaction} \label{basic:sphere-s}
The outermost end of the tip is the most important part of the probe
as it is closest to the sample surface. Owing to its
dimensions and proximity to the sample surface, this part of the
probe experiences the strongest interaction with the surface and
accounts for the achievable lateral resolution. In a first
approximation it can be described by a sphere of radius~$r$. Typical
values for~$r$ range from 10~to~100\,nm.
As a direct consequence of the small tip radius, the charge density
at its surface is very high, resulting in a strong electrostatic
force between sphere and surface charges which can lead to an
indentation~$h_{\rm el}$ of the tip into the surface ('Hertzian
indentation'). A detailed theoretical analysis of this mechanism was carried out by Kalinin and Bonnell~\cite{Kal02} leading to:
\begin{equation}\label{eq:joh-01}
    h_{\rm el} = \frac{2}{3}\left( \frac{3}{4E^{\ast}} \right)^{2/3} r^{-1/3} F^{-1/3} F_{\rm el}
\end{equation}
with~$E^{\ast}$ being the effective Young's modulus of the tip-surface system. $E^{\ast}$ can be obtained from the relation $\frac{1}{E^{\ast}} =  \frac{1-\nu_1^2}{E_1} + \frac{1-\nu_2^2}{E_2}$ where $E_1$, $E_2$ and $\nu_1$, $\nu_2$ are Young's moduli and Poisson ratios of tip and surface materials. $F$~denotes the load of the tip and $F_{\rm el}$~the electrostatic contribution to the load.

The main part of the tip consists of a cone with an opening angle of
typically $\alpha = 30^{\circ}$ and a height of 10 to 20\,\textmu m.
Any SFM-imaging owing to an interaction between the cone and the sample surface would thus be limited to a lateral resolution of $> 10$\,\textmu m. A contribution of the cone can therefore only act as a background. In standard SFMs the cone is inclined by
$\beta = 20$ -- 30$^{\circ}$ relative to the normal of the surface.
A close-up view of the situation is shown in
Fig.~\ref{fig:Joh01}(b).
Owing to the asymmetry of the electric field caused by the
inclination~$\beta$ one could think of a net force acting at
the edge of the cone which is closest to the surface. This would
lead to a change of the inclination of the cone with the outermost
end of the tip as pivot and thus a deformation of the cantilever.

\subsection{Cantilever-Sample Interaction} \label{basic:cantilever-s}
The tip is attached to the cantilever whose typical dimensions are
of the order of 100 to 500\,\textmu m for the length and 25 to
50\,\textmu m for the width. The minimum distance between the
cantilever and the sample surface is given by the tip height (10 to
20\,\textmu m). Typically, the cantilever is inclined by $\beta =
20$ -- 30$^{\circ}$ relative to the surface (Fig.~\ref{fig:Joh01}(a)). These numbers alone
show the limits of SFM-imaging owing to an interaction between the cantilever and the sample surface, and again, only a background signal can be expected from this interaction.
When imaging surfaces with strong topographic features, this argumentation does not hold any longer since in this case, the overall average sample-cantilever spacing would change, and thus the measured SFM-signal. Because the roughness of our samples is $\ll$5\,nm only, and the tip height is $>10$\,\textmu m, we expect this mechanism to play a minimal role.

\section{Alternative Model}\label{sec:model}
\begin{figure*}[ppp]
\includegraphics{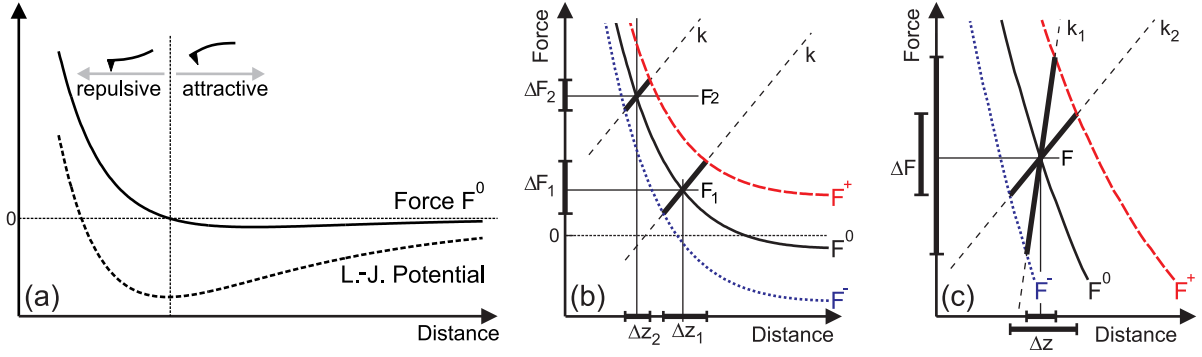}
\caption{\label{fig:Joh02}
(a) Lennard-Jones potential and its derivative, the force-distance curve.
(b and c) Force-distance curve ($F^0$) within the repulsive part of the Lennard-Jones potential. An additional electrostatic force shifts this curve upwards ($F^+$-\,-\,-) or downwards ($F^-\cdots$). Applying an oscillating voltage to the tip leads to force changes $\Delta F$ followed by distance changes $\Delta z$.
In (b) the situation for two different setpoints of the feedback ($F_1 < F_2$) using the same cantilever (spring constant $k$) is shown. In (c) a close-up view shows the effect of
the spring constant ($k_1 > k_2$) on the force and distance changes
$\Delta F$ and $\Delta z$ respectively for the same feedback
setpoint.
}
\end{figure*}

In addition to the basic considerations described above there is another contrast mechanism for electrostatic interactions in contact-mode SFM which we will explore in the following. The tip-sample interaction is generally described by a modified Lennard-Jones potential
\begin{equation}\label{eq:Joh99}
W(z) = -\frac 23\pi^2\varepsilon\rho_1\rho_2\sigma^5r\left[\frac
\sigma z - \frac 1{210} \left(\frac{\sigma}{z}\right)^7\right]
\end{equation}
with $z$ being the tip-sample distance, $\varepsilon$~the depth of
the potential, $\rho_1$~and $\rho_2$~the number densities of tip and sample respectively, $r$~the tip radius and $\sigma$~the equilibrium distance~\cite{Sarid,Mey89}. The force between tip and sample surface equals minus the derivative of the potential, thus (Fig.~\ref{fig:Joh02}(a)):
\begin{equation}\label{eq:Joh00}
F^0(z) = - \frac
23\pi^2\varepsilon\rho_1\rho_2\sigma^4r\left[\left(\frac
{\sigma}{z}\right)^2 - \frac 1{30}
\left(\frac{\sigma}{z}\right)^8\right]
\end{equation}
%


 The graphs shown in Fig.~\ref{fig:Joh02}(b and c) are close-up views within the repulsive part of the Lennard-Jones potential. The setpoint of the feedback is seen as the intersection point of the force-distance curve~$F^0(z)$ and the straight dotted lines corresponding to the spring constant~$k$.
In the case of an additional electrostatic force between tip and
sample surface this force-distance curve is shifted up- or downwards
($F^+$ or $F^-$), depending on the relative polarity between tip and
surface charge.
If an alternating voltage $\widetilde{U}=U_0\sin \omega t$ is applied to the tip, the setpoint of the feedback will oscillate between two positions as indicated by the thick bars in Fig.~\ref{fig:Joh02}(b and c). Since the frequency $\omega$ is high with respect to the time constant of the SFM-feedback, the tip oscillates along one of those bars, thereby experiencing force changes $\Delta F$ followed by distance changes $\Delta z$. The latter can be read out as the SFM$_{\rm el}$-signal with a lock-in amplifier.

For a better understanding of the contrast mechanism described
above, we will specifically address the situation for different  setpoints as well as various spring constants.
Figure~\ref{fig:Joh02}(b) shows the situation for two different
feedback  setpoints of the SFM, i.\,e.~two different loads of the tip ($F_1<F_2$) using a cantilever with spring constant $k$. As is obvious from the graph, the higher the load, the smaller the distance change ($\Delta z_2 < \Delta z_1$). The effect of the stiffness of the cantilever when using the same setpoint can be seen in Fig.~\ref{fig:Joh02}(c): the cantilever with the higher spring constant ($k_1$) will result in a lesser response.

The properties of the proposed contrast mechanism can be summarized
as follows: (i) it is independent of any material parameters and
(ii) soft cantilevers show a greater response.


%
\section{Experimental} \label{sec:exp}
The experiments were carried out with a commercial scanning force
microscope (SMENA from NT-MDT), modified to allow the application of voltages to the tip. To perform long-range scanlines we upgraded the
sample-holder with a piezo-sustained translation stage. This allowed
us to reproducibly displace the sample by~250\,\textmu m. For the experiments we used two operation modes of the SFM: contact mode and non-contact mode~\cite{MHB,Sarid}. In contact mode, the tip touches the surface and is located in the repulsive part of the Lennard-Jones potential (Fig.~\ref{fig:Joh02}(a)). In non-contact mode, the tip never touches the sample surface; it is thus situated in the attractive part of the Lennard-Jones potential. Irrespective of the operation mode, an additional voltage can be applied to the conductive tip to map both the charge distribution and/or the piezoelectric response of the sample. To improve the sensitivity, this voltage is applied as an alternating voltage $\widetilde{U}=U_0\sin\omega t$ and the cantilevers vibration can thus be read out as in-phase signal (X-channel) using a lock-in amplifier (SRS 830; Stanford Research Systems). Typical frequencies are of the order of $\omega = 10$--100\,kHz and the amplitude is usually $U_0<15$\,V$_{\rm pp}$. Since this frequency~$\omega$ can be chosen freely, it is advisable to set it far from the mechanical resonance frequency of the cantilever~$\omega_{\rm n}$ to avoid any mutual influence.

\begin{figure*}[ppp]
\includegraphics{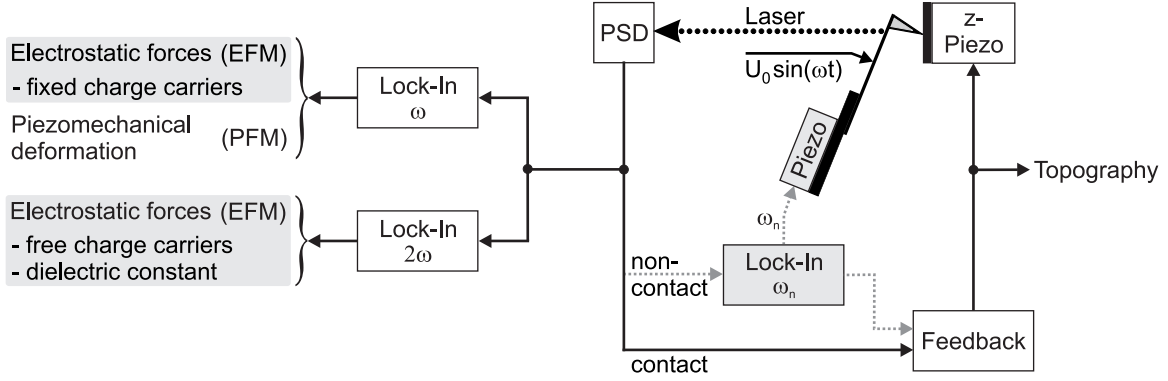}
\caption{\label{fig:Joh03}
The right side shows a schematic of the electric circuit for both, the contact and the non-contact mode. For the non-contact mode, the tip is excited at its resonance frequency $\omega_{\rm n}$ and its vibration amplitude is read out with a lock-in amplifier operated at $\omega_n$. Although depicted in the same schematic, the two  operation-modes can, of course, not be run simultaneously.
To allow for electrostatic force microscopy (EFM) and piezoresponse force microscopy (PFM) the tip is electrically connected to an alternating voltage $\widetilde{U}=U_0\sin\omega t$. The two lock-in amplifiers on the left side, operating at frequencies $\omega$ and $2\omega$ are utilized for readout of the electrostatic interactions between tip and surface charges (according to Eq.~\ref{fig:Joh07}) and the piezoresponse of the sample. Note that irrespective of the operation mode chosen, the output signals from these two lock-in amplifiers can be recorded.
PSD: position sensitive detector.
}
\end{figure*}

In the following we will recall the specific properties of
piezoresponse force microscopy and electrostatic force microscopy (Fig.~\ref{fig:Joh03}).

\subsection{Piezoresponse Force Microscopy (PFM)} \label{exp:PFM}
To determine the piezomechanical response of the sample the SFM
is operated in contact mode with an alternating voltage
$\widetilde{U}=U_0\sin\omega t$ applied to the tip~\cite{Alexe}. This leads to periodic thickness changes of the sample underneath the tip via the converse piezoelectric effect which can be read-out with a lock-in amplifier. The thickness change~$\Delta t$ is
\begin{equation}\label{eq:Joh04}
\Delta t = U\,d
\end{equation}
with~$U$ being the applied voltage and $d$ the
appropriate piezoelectric coefficient. Note that (i) the thickness
change~$\Delta t$ is independent of the thickness~$t$ of the sample
itself and (ii) the inhomogeneous electric field distribution from
the tip does not lead to an enhancement of the deformation as
$\int_0^t E \rm{d} s = U$~\cite{Jun06}.
This technique is mainly used for mapping ferroelectric domain
patterns as ferroelectricity entails piezoelectricity. The
detection method is very sensitive: it allows  for mapping ferroelectric domains in materials with piezoelectric coefficients $< 0.1$\,pm/V thus leading to thickness changes of 1\,pm only when applying 10\,V$_{\rm pp}$ to the tip. The lateral resolution of PFM is given by the tip radius and limited to $\approx 10$\,nm~\cite{Jun08}.

\subsection{Electrostatic Force Microscopy (EFM)} \label{exp:EFM}
Charges~$q$ at the surface are generally detected by use of the SFM
in non-contact mode with a voltage $U$
applied to the tip~\cite{Ste88}. The tip is thus charged by
\begin{equation}\label{eq:Joh01}
Q_{\rm tot} = q_{\rm i} + Q
\end{equation}
with the image charge~$q_{\rm i}$ and the capacitive charging $Q =
C\, U$, whereby $C$ is the capacity of the tip-sample system. When applying an alternating voltage $\widetilde{U}=U_0\sin\omega t$ the tip senses the electrostatic force
\begin{equation}\label{eq:Joh02}
      F_{\rm el} = \frac{q\,Q_{\rm tot}}{4\pi \varepsilon_0 z^2}
        + \frac{1}{2}\frac{{\rm{d}} C}{{\rm{d}} z} (U_0\sin\,\omega t)^2
\end{equation}
with the tip-sample distance~$z$. Expanding Eq.~\ref{eq:Joh02}
results in the frequency sorted form:
\begin{eqnarray}\label{eq:Joh03}
  \nonumber
  F_{\rm el}&=&\frac{q\,q_{\rm i}}{4\pi\varepsilon_0 z^2}
       \,+\,\frac{1}{4}\frac{{\rm{d}} C}{{\rm{d}} z} U_0^2 \\
  &+& \,\frac{q\,C}{4\pi\varepsilon_0 z^2}U_0   \sin(\omega t)
       \,-\,\frac{1}{4}\frac{{\rm{d}} C}{{\rm{d}} z}U_0^2   \cos(2\omega t)
\end{eqnarray}
The first term denotes the Coulomb force between the surface charge
$q$ to be measured and its image charge~$q_{\rm i}$ on the tip and
is independent of the voltage applied to the tip.
The second term in Eq.~\ref{eq:Joh03} describes the static part of
the capacitive force between tip and sample surface and is
independent of the surface charge $q$. The third term is the one
that governs the charge detection, as it is proportional to the
charge~$q$ to be measured and the charges generated on the
tip by the voltage $\widetilde{U}=U_0\sin \omega t$. Finally, the last term comprises both the free charge carrier density~\cite{charges} and the polarizability (i.\,e.~the dielectric constant) of the sample~\cite{Mar88}.
Therefore, recording simultaneously the cantilever oscillations at
the frequencies~$\omega$ and~$2 \omega$ allows one to map the distribution of the fixed charges (i.e.~the electrostatic field distribution) as well as the compound response of free charge carriers and polarizability. In Sec.~\ref{res:c-nc} where both signals are discussed, they will be explicitly named SFM$_{\rm el}(\omega)$ and SFM$_{\rm el}(2\omega)$. If not specified SFM$_{\rm el}(\omega) = {\rm SFM}_{\rm el}$.


\subsection{Samples and Probes} \label{exp:sampprob}
\begin{figure}[ppp]
\includegraphics{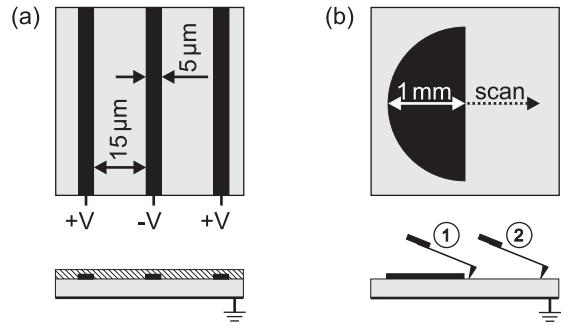}
\caption{\label{fig:Joh04}
(a) \textmu m-sized Au-stripes evaporated on top of a glass plate
and covered with a 400\,nm thick dielectric layer. (b) Evaporated
large area gold-electrode. In this case no dielectric layer was
deposited on top. For the measurements we performed scans between
the two positions shown in the side-view.
}
\end{figure}

The experiments for the interaction between tip and surface charges
require samples where a (localized) electrostatic interaction can be
generated as and when required. We therefore prepared a series of
samples where we evaporated \textmu m-sized gold-structures of
50\,nm thickness onto a glass slide. These Au-stripes could be
individually electrically connected (Fig.~\ref{fig:Joh04}(a)).
We then spin-coated varnishes of different hardness on top of it:
 alkali silicate (AS) and   photoresist (PR).  The materials' parameters are as follows: poisson ratios $\nu_{\rm AS}= 0.23$ and $\nu_{\rm PR}= 0.37$, Young's moduli $E_{\rm AS}= 70\,$GPa and $E_{\rm PR}= 3.7\,$GPa~\cite{Oli92,Cro00}. The thickness of the varnishes was 400\,nm each. This could be determined by SFM after we scratched the surface with a diamond tip.
The varnish also prevented short-circuiting between the tip and the
conductive Au-structures when the SFM was operated in contact mode.


To determine the interaction between the cantilever and surface
charges we evaporated a large-area (some mm$^2$) Au-structure with a
sharp edge, again on top of a glass slide (Fig.~\ref{fig:Joh04}(b)).
This sample was not covered with a dielectric varnish.
There are two extreme positions of the cantilever with respect to
the Au-electrode: \textcircled{1} the full length of the cantilever
is above the electrode and \textcircled{2} the cantilever is fully
beside the electrode. To perform a scan between those two positions
we utilized the piezo-sustained translation stage.

We used a series of different probes listed in Tab.~\ref{tab:Joh01}. As for the coatings, we would like to point out that all probes are made out of highly $n$-doped silicon, and therefore the material itself is conductive. Owing, however, to the oxidation of silicon, uncoated probes are mostly instantaneously covered by a thin layer (a few nm) of non-conductive~$\rm SiO_2$.

For the measurements of the impact of the tip-sample interaction, we
always oriented the sample in such a way that the direction of the
cantilever was perpendicular with respect to the Au-stripes
(Fig.~\ref{fig:Joh04}(a)), thus minimizing the influence of the
cantilever on the measurements. When investigating the interaction
between the cantilever and the sample, we controllably positioned
the cantilever above the Au-area (Fig.~\ref{fig:Joh04}(b)).

\section{Results and Discussion} \label{sec:results}

In this section we will present our experimental results and discuss the origin of a possible $\rm SFM_{el}$-signal in contact mode operation. We will separately focus on the results of the tip-sample interaction, potentially enabling laterally resolved imaging (Sec.~\ref{res:sphere-s}). We then shortly present results on the cantilever-sample interaction (Sec.~\ref{res:cantilever-s}). Finally we discuss comparative measurements of the $\rm SFM_{el}$-signal using both, contact and non-contact mode operation of the SFM (Sec.~\ref{res:c-nc}).

\subsection{Tip-Sample Interaction} \label{res:sphere-s}
In a first series of experiments we investigated the influence of the hardness of the sample surface on the SFM$_{\rm el}$-signal, thereby testing the applicability of the Hertzian indentation model  (Sec.~\ref{basic:sphere-s}) for the present situation. The SFM is operated in contact mode and we used samples such as the one depicted in Fig.~\ref{fig:Joh04}(a) which were covered with soft photoresist  and hard alkali silicate. We recorded the SFM$_{\rm el}$-signal (see Eq.~\ref{eq:Joh03}) while applying a DC-voltage of $10$\,V to one of the Au-stripes, the tip being positioned on top of it. Interestingly we could not observe any difference in the amplitude of the SFM$_{\rm el}$-signal measured with both samples, although the material properties differ strongly, presumably leading to a difference for $h_{\rm el}$ and thus SFM$_{\rm el}$ by one order of magnitude according to Eq.~\ref{eq:joh-01}.

\begin{figure}[ppp]
\includegraphics{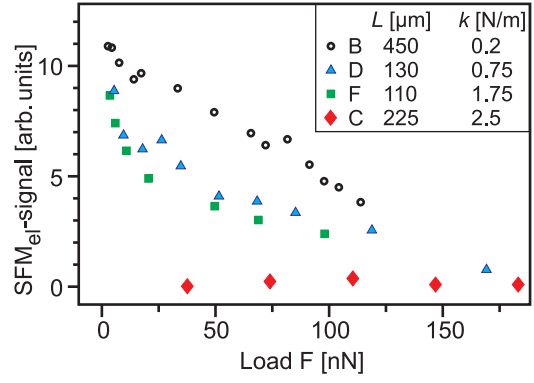}
\caption{\label{fig:Joh05}
Dependence of the $\rm SFM_{el}$-signal on the load $F$ of the tip using  different probes ($L$: length, $k$:
spring constant of the cantilever) measured with the sample shown in
Fig.~\ref{fig:Joh03}(c).    The voltage applied to the Au-electrode
was 10\,V. The different probes are listed in Tab.~\ref{tab:Joh01}.
}
\end{figure}

In a second series of experiments we investigated the influence of
the load~$F$ of the tip on the amplitude of the $\rm
SFM_{el}$-signal, the SFM again being operated in contact mode. We therefore positioned the tip above one of the Au-stripes (again with a DC-voltage of 10\,V applied) and measured the $\rm SFM_{el}$-signal for different loads~$F$. We repeated the experiment with four different probes and thus four different cantilever spring constants~$k$. The results are shown in Fig.~\ref{fig:Joh05}. According to Eq.~\ref{eq:joh-01} only the load~$F$, but not the spring constant~$k$ of the cantilever used, should influence $h_{\rm el}$ following a $F^{-1/3}$-dependence. Our measurements, however, do not show such a dependence and obviously the smaller the spring constant the larger the $\rm SFM_{\rm el}$-signal for a given load~$F$.

From the results given above, we concluded that Hertzian indentation is at least not the dominant mechanism responsible for the $\rm SFM_{el}$-signal in contact-mode operation. On the other hand, the observed dependencies agree with the expectations from the proposed alternative model described in Sec.~\ref{sec:model}.

In order to learn about a possible contribution of the cone to an
electrostatic interaction between tip and sample, we recorded images
on periodically poled lithium niobate, exhibiting a very strong
domain-specific surface charging of $\rm 0.7\,C/m^2$. Changing the angle $\beta$ (Fig.~\ref{fig:Joh01}) from $0^{\circ}$ to $80^{\circ}$ did not result in any change of the recorded signal at frequency $\omega$. We therefore concluded that any contribution of the cone to the SFM$_{\rm el}$-signal owing to its inclination with respect to the sample surface is not relevant.

\subsection{Cantilever-Sample Interaction} \label{res:cantilever-s}

\begin{figure}[ttt]
\includegraphics{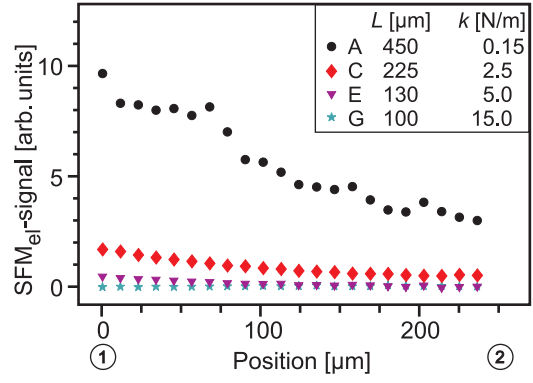}
\caption{\label{fig:Joh06}
Measurement of the electrostatic force between the cantilever
and a homogeneous electrode for different probes ($L$: length, $k$:
spring constant of the cantilever) measured with the sample shown in
Fig.~\ref{fig:Joh04}(b).    The voltage applied to the Au-electrode
was 10\,V. The different probes are listed in Tab.~\ref{tab:Joh01}.
}
\end{figure}

To complete the picture of electrostatic interaction between probe
and sample when the SFM is operated in contact mode, we performed a series of experiments to measure the contribution of the cantilever to the SFM$_{\rm el}$-signal using the sample depicted in Fig.~\ref{fig:Joh04}(b). We therefore determined the dependence of the SFM$_{\rm el}$-signal on the overlap between cantilever and Au-electrode between the two positions \textcircled{1} and \textcircled{2} indicated in Fig~\ref{fig:Joh04}(b).
As it is obvious from Fig.~\ref{fig:Joh06} the probe with the softest cantilever shows the largest signal, but because its length is $L=450$\,\textmu m and the displacement is limited to 250\,\textmu m we could only record part of the sloping curve. The amplitudes of the SFM$_{\rm el}$-signal of the four cantilevers at position \textcircled{1} directly reflect the spring constants and the lengths respectively. These results perfectly meet the expectations according common sense.

We would like, however, to emphasize that although clearly measurable, the SFM$_{\rm el}$-signal on flat surfaces caused by the can\-ti\-le\-ver-sample interaction can only account for a possible background because of the vast size of the cantilever of at least $30 \times 100$\,\textmu m$^2$. Consequently any features in the recorded SFM images smaller than the area of the cantilever are due to an interaction between the tip and the sample, namely the sphere (since we excluded a contribution of the cone).

\subsection{Comparison contact - non-contact mode operation}\label{res:c-nc}


Finally for the sake of completeness we show a series of images comparing contact and non-contact mode recording of SFM$_{\rm el}$-signals. We therefore simultaneously measured the SFM$_{\rm el}(\omega)$- and the SFM$_{\rm el}(2\omega)$-signal with the sample depicted in Fig.~\ref{fig:Joh04}(a) and the electrical set-up
shown in Fig.~\ref{fig:Joh03}.

\begin{figure*}[ttt]
\includegraphics{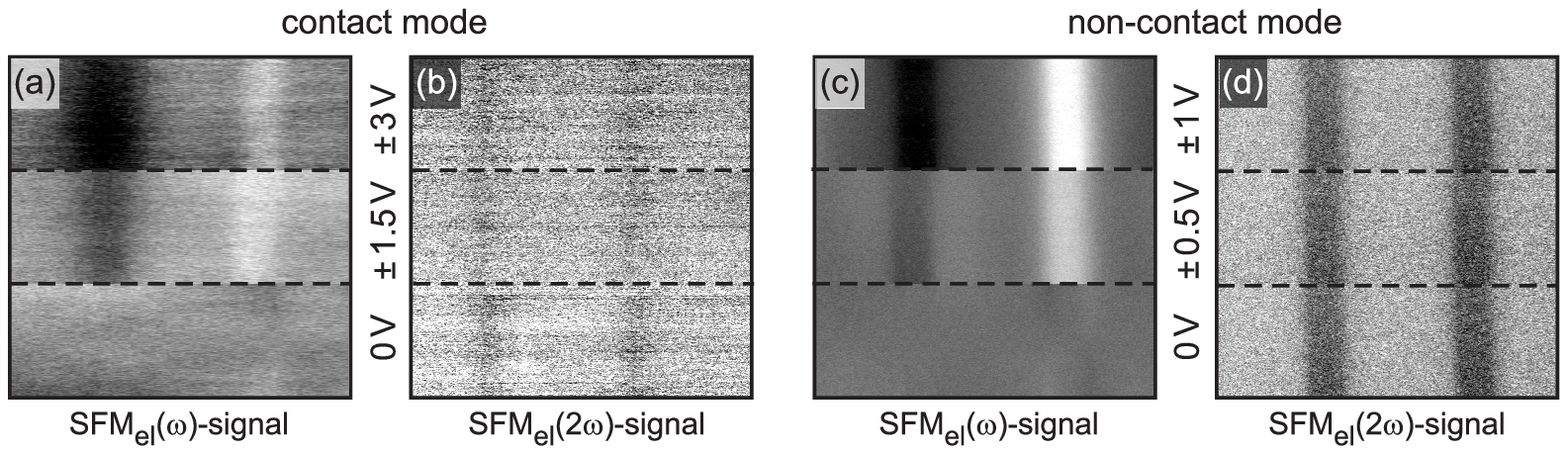}
\caption{\label{fig:Joh07}
Electrostatic forces measured in contact mode (a, b) and in non-contact mode (c, d)  using the sample from Fig.~\ref{fig:Joh04}(a). During image acquisition (after 1/3 and again after 2/3), we changed the DC-voltages applied to the Au-stripes as indicated in-between the two images, positive for the left and negative for the right stripe, respectively.  For the contact mode recording a soft cantilever was used with 42\,V$_{\rm pp}$ applied to the tip. For the non-contact imaging we used a harder cantilever and 14$\,$V$_{\rm pp}$. Image size is 45$\times$45\,\textmu m$^2$.
}
\end{figure*}

As can be seen in Fig.~\ref{fig:Joh07}, both operation modes allow us to detect the fixed charge distribution (a and c) and the  free charge carrier density (b and d).
Whereas for the SFM$_{\rm el}(\omega)$-signal the observed contrast depends on the polarity as well as the magnitude of the DC-voltage applied to the stripes this is not the case for the SFM$_{\rm el}(2\omega)$-signal. It is furthermore obvious that the contrast, and together with that the lateral resolution, is enhanced in non-contact mode compared with the contact mode. The scanning speed was the same for both operation modes as well as the settings of the lock-in amplifiers (time constant $\tau$:~3\,ms; sensitivity $S$:~500\,\textmu V). The amplitude of the alternating voltage applied to the tip was 42\,V$_{\mathrm{pp}}$ for the contact mode and 14\,V$_{\mathrm{pp}}$ for the non-contact mode, the cantilevers used had spring constants of 0.2\,N/m and 5\,N/m, respectively.

The unsatisfying situation that cantilevers with different spring constants were used for the contact and the non-contact mode measurements are justified as follows: on the one hand hard cantilevers $k \ge 2$\,N/m, showed no SFM$_{\rm el}(\omega)$-signal in contact-mode operation, while on the other hand soft cantilevers do not allow for non-contact mode operation. As it seems there is, unfortunately, no reasonable overlap, i.\,e., a cantilever allowing for both  recording a SFM$_{\rm el}(\omega)$-signal in contact mode and enabling non-contact mode operation.

The results shown in Fig.~\ref{fig:Joh07} agree favorably with the expectations described before (Sec.~\ref{sec:exp}): the SFM$_{\rm el}(\omega)$-signal, i.\,e.~the third term in Eq.~\ref{eq:Joh03}, depicting the fixed charge carriers~$q$ depends on the voltage applied to the Au-stripes. In terms of electrostatic interaction a fixed charge carrier or a voltage applied to a locally limited conductor both result in an electric field and can thus be measured as a SFM$_{\rm el}(\omega)$-signal. The SFM$_{\rm el}(2\omega)$-signal, i.\,e.~the fourth term in Eq.~\ref{eq:Joh03}, only reflects the material properties of gold in comparison with the surrounding insulators, thereby being independent on the voltage applied to the Au-stripes. And finally, only soft cantilevers allow for recording a SFM$_{\rm el}(\omega)$-signal in contact mode operation, in accordance with our model described in Sect.~\ref{sec:model}. The fact that in non-contact mode where the tip can move freely, the $\rm SFM_{el}(\omega)$-signals are larger (in our measurement by a factor of 60) than in contact mode where the tip presses against the surface, is self evident and therefore needs not further comment.

\subsection{Quantitative estimate on the electrostatic contribution to the PFM signal}\label{res:PFM-EFM}

Finally we want to come back to our starting point: do surface polarization charges of ferroelectric materials influence PFM measurements? We therefore firstly compared the electrostatic force acting on the probe at large tip-sample distance above a single domain lithium niobate crystal~\cite{Joh09} to that from the sample depicted in Fig.~\ref{fig:Joh04}(b). $\rm LiNbO_3$ is somewhat an extreme material as it has on the one hand a very high surface charge density (0.7\,C/m$^2$) but on the other hand a very small piezoelectric coefficient of only $d_{33}\approx\rm6\,pm/V$. It turned out that applying $\approx150$\,V to the Au-electrode led to similar SFM$_{\rm el}$-signals at large distance. Next we compared the amplitude of the PFM-signal on the LiNbO$_3$ crystal with that of the SFM$_{\rm el}$-signal obtained on the Au-electrode (applying 150\,V to it), the SFM being operated in contact mode. Our measurements revealed that the amplitude of the SFM$_{\rm el}$ signal is $\ll 5$\% when compared to that of the PFM-signal.

\section*{Conclusions}\label{sec:conclusions}
In this contribution we have investigated the
possibilities of recording charge distributions using contact mode
scanning force microscopy (SFM).
Utilizing especially designed samples we could show that Hertzian
indentation has no significant impact on the observed SFM signals,  but rather the altered conditions of the tip-sample interaction owing to the electrostatic forces between tip and surface charges, which in turn affect the setpoint of the scanning force microscope, lead to an image contrast of charge distributions.
Hence under certain conditions, it is possible to detect
charge distributions in contact-mode SFM. The use of very soft cantilevers (spring constants $k < $1\,N/m), however, is mandatory for obtaining an image contrast.

The starting point of our investigations was the question about a possible electrostatic contribution to the signals recorded with piezoresponse force microscopy (PFM) on ferroelectric surfaces. Based on our experimental results the electrostatic interaction between the charged tip and the polarization charges at the sample surface can basically be neglected in PFM measurements. Moreover the fact that for PFM one generally uses hard cantilevers (spring constants $k > $2\,N/m) further more reduces any electrostatic contribution to the PFM signal.

{\small {\bfseries Acknowledgments}
We thank Tobias Jungk for fruitful discussions. Financial support
from the Deutsche Telekom AG is gratefully acknowledged.
}

\eject

\eject

\begin{table*}[ppp]
\caption{\label{tab:Joh01}
Probes used for the experiments. The values given for the dimensions
are the nominal values from the manufacturers. The spring constant
was determined via the measured resonance frequency. The symbols
correspond to those used in Figs~\ref{fig:Joh05} and \ref{fig:Joh06}.
\newline Manufacturers: $^{(1)}$Nanosensors; $^{(2)}$MikroMasch; $^{(3)}$NT-MDT
}
\vspace{1cm}
\begin{tabular}
{|c|c|c|c|c|c|c|}
  \hline
   Label  & Length & Width  & Tip Radius &Spring Constant  & Coating & Type   \\
            & $L$ [\textmu m] & $W$ [\textmu m] &$r$ [nm] & $k$ [N/m] &  &  \\ \hline
  $\bullet$\quad A  & 450 & 48 & 10 & 0.15   & none & PP-S3G3T1-3$^{(1)}$ \\
  $\includegraphics{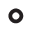}$\quad B  & 450 & 48 & 10 & 0.2\phantom{0}    & none & PP-S3G3T1-3$^{(1)}$ \\
  $\blacklozenge$\quad C  & 225 & 28 & 10 & 2.5\phantom{0}    & none & PP-FM$^{(1)}$ \\
  $\blacktriangle$\quad D & 130 & 35 & 40 & 0.75 & Ti-Pt & NSC36$^{(2)}$ \\
  $\blacktriangledown$\quad E & 130 & 35 & 70 & 5.0\phantom{0} & Diamond & DCP11$^{(3)}$ \\
  $\blacksquare$\quad F& 110       & 35 & 40 & 1.75   & Ti-Pt & NSC36$^{(2)}$ \\
  $\bigstar$\quad G & 100     & 35 & 70 & 15.0\phantom{00}   & Diamond & DCP11$^{(3)}$ \\
  \hline
\end{tabular}
\end{table*}

\clearpage

\clearpage

\clearpage

\clearpage

\clearpage

\clearpage

\clearpage

\begin{thebibliography}{}
\bibitem{charges}
The visibility of the free charge carries density was demonstrated using a photosensitive sample and recording the dependence of the  $2\omega$ response on the light irradiating power.
\end{thebibliography}
\end{document}